\documentclass[fleqn,10pt]{wlscirep}
\usepackage[utf8]{inputenc}
\usepackage[T1]{fontenc}

\usepackage{graphicx}
\usepackage{subcaption}
\usepackage{placeins}
\usepackage{verbatim}

\usepackage{graphicx,epstopdf}
\usepackage[utf8]{inputenc}
\usepackage[T1]{fontenc}
\usepackage{chemformula}

\usepackage{adjustbox}
\usepackage{lipsum}
\usepackage{mathtools}
\usepackage{physics}
\usepackage{floatrow}
\usepackage{hyperref}
\usepackage[justification=centering]{caption}
\usepackage{amsmath}
\usepackage{nccmath}
\usepackage{floatrow}
\usepackage[thinlines]{easytable}
\usepackage{cleveref}
\usepackage{gensymb}
\usepackage{booktabs}  
\usepackage{ltablex}
\usepackage{bbm}
 
\usepackage{subcaption} 
 
\usepackage{soul}
\usepackage{cancel}

\usepackage{chemformula}

\title{Optical polarization evolution and transmission in multi-Ranvier-node axonal myelin-sheath waveguides}
\author[1,2,3]{Emily Frede}
\author[1,2,3,*]{Hadi Zadeh-Haghighi}
\author[1,2,3,*]{Christoph Simon}
\affil[1]{Department of Physics and Astronomy, University of Calgary, Calgary, AB T2N 1N4, Canada}
\affil[2]{Institute for Quantum Science and Technology, University of Calgary, Calgary, AB T2N 1N4, Canada}
\affil[3]{Hotchkiss Brain Institute, University of Calgary, Calgary, AB T2N 1N4, Canada}

\affil[*]{hadi.zadehhaghighi@ucalgary.ca, csimo@ucalgary.ca}

\begin{abstract}
In neuroscience, it is of interest to consider all possible modes of information transfer between neurons in order to fully understand processing in the brain. It has been suggested that photonic communication may be possible along axonal connections, especially through the myelin sheath as a waveguide, due to its high refractive index. There is already a good deal of theoretical and experimental evidence for light guidance in the myelin sheath; however, the question of how the polarization of light is transmitted remains largely unexplored. It is presently unclear whether polarization-encoded information could be preserved within the myelin sheath. We simulate guided mode propagation through a myelinated axon structure with multiple Ranvier nodes. This allows both to observe polarization change and to test the assumption of exponentiated transmission loss through multiple Ranvier nodes for guided light in myelin sheath waveguides. We find that the polarization can be well preserved through multiple nodes and that transmission losses through multiple nodes are approximately multiplicative. These results provide an important context for the hypothesis of neural information transmission facilitated by biophotons, strengthening the possibility of both classical and quantum photonic communication within the brain. 
\end{abstract}
\begin{document}

\flushbottom
\maketitle
%
%
\thispagestyle{empty}

\section*{Introduction} 
The brain is a complex system of billions of neurons, which are structurally interconnected by axons. It is well known these axonal connections facilitate electrochemical signals between neurons in the form of action potentials. However, many unanswered questions remain in neuroscience\cite{neuroq-general}, notably with regards to mechanisms of higher-order functions such as learning\cite{neuroq-learning}, memory\cite{neuroq-memoryold, neuroq-memorynew}, and consciousness\cite{neuroq-consciousness, neuroq-anesthesia, neuroq-consciousness-and-anes}. Difficulties in understanding these higher-order functions may indicate that electrochemical signals are not the only modality of information processing in the brain. It has been proposed that additional information transfer may occur through the exchange of biophotons between neurons\cite{biophoton-neural-signals, biophoton-brain-signal-review, simulated-biophotons-cause-transsynaptic-signals}, especially along myelinated axons as waveguides\cite{Sourabh, group-review, biophoton-propagation-simulation} due to the high refractive index of myelin\cite{myelin-refractive-index}. \\
\\
Biophotons are produced within the wavelength range of $350 \: \text{nm}$ -- $1300 \: \text{nm}$ by oxidative processes in living cells\cite{UWPE-review}. Their estimated production rate is 1 - 10,000 photons per second per centimeter of tissue squared\cite{UWPE-review} -- though this rate might be higher inside cells\cite{biophoton-high-rate}. They have been observed in a range of organisms and tissues\cite{UWPE-review}, and their spectral range is affected by species\cite{species-redshift} and the aging process\cite{aging-mouse-brain}. Their existence in the brain is well-established\cite{rat-hippocampus-slices, in-vivo-activity-imaging, neural-tissue-activity-imaging, rat-eyes-imaging, spatiotemporal-glutamate-imaging, UWPE-review}, and their exchange between neurons is a potential means of information transfer. Photons may play a role in cellular communication generally\cite{fels-communication}. It is argued that biophoton production is a good candidate for encoding information\cite{biophoton-brain-signal-review}, as their originating oxidative processes also produce reactive oxygen species known to be strongly regulated cellular signals in biological systems\cite{ROS-on-synapses,ROS-in-physiology,ROS-functional-roles,ROS-in-signalling}. In an \textit{in vivo} examination of rat brain, biophoton emission intensity was correlated with neural activity through electroencephalography observations \cite{in-vivo-activity-imaging}, which indicates emission may be biologically meaningful. Photons serving as information carriers could have many functions within neural information processing, e.g., facilitating backpropagation, a well-known mechanism for learning, within the brain\cite{Parisa}. \\
\\
Myelinated axons are the predominant nerve fibers found in white matter, which makes up about half of the human brain\cite{white-matter-percentage}. On such axons, the myelin sheath envelopes the axon as an insulating lipid layer which increases the propagation speed of action potential signals\cite{myelin-speeds-up-action-potentials}. The myelin sheath, therefore, inherits bends and cross-sectional shape from the axon and is interrupted at regular intervals by nodes of Ranvier which are vital for its function in speeding up action potentials\cite{node-length-affects-signal-speed}. In the central nervous system, myelin is formed by the oligodendrocyte glial cells. Light guidance in the myelin sheath would not be a new function for glial cells, as Müller cells are known to guide light in mammalian eyes\cite{muller-cell-optical-fibers, muller-cells-separate-wavelengths}. It is also known that the myelin sheath is not a static structure, which naturally affects its properties as a waveguide. Myelination is a plastic process, such that axons become myelinated and myelin sheaths can undergo structural changes (e.g., changes in thickness, internodal lengths, Ranvier node geometry, etc.) throughout the entire lifetime\cite{myelin-plasticity}. \\
\\
The first study simulating myelin sheath waveguides was conducted by Kumar, et al. in 2016\cite{Sourabh}, which assessed whether myelinated axons could serve as biophoton waveguides despite realistic optical imperfections. They simulated (biophoton spectral range) mode propagation through models of myelinated axon segments, informed by biologically-observed optical properties, finding transmitted light well-confined to the myelin sheath. Their models assessed light transmission in the context of imperfections that can interfere with light-guiding ability, most notably a Ranvier node, bends in the axon, varying myelin sheath thickness, and non-circular nerve fiber cross-sections. Their treatment to estimate transmission across inter-neuron distances demonstrated that biophoton transmission along myelinated axons is possible within realistic expectations of these optical imperfections. This has sparked various theoretical studies\cite{biophoton-propagation-simulation, ARROW-nerve-fiber-simulation, analytical-myelinated-axon, bio-nanoantennas, infrared-simulation, infrared-in-vivo-study, engineering-for-demyelination}, and the possibility of myelin sheath waveguides has been well-supported through subsequent simulations. Another biophoton-motivated simulation study\cite{biophoton-propagation-simulation} considered (optical spectral range) mode propagation in the context of a myelinated axon model containing the detailed multi-lamellar structure of the myelin sheath. They found an optimum of low attenuation and low dispersion within realistic parameters of this model for light operating wavelengths in the biophoton wavelength range. Dependence in operating wavelength due to axon diameter and myelin thickness, especially as may be implemented by myelination plasticity, might explain spectral redshift of biophotons in species\cite{species-redshift} and blueshift during the aging process\cite{aging-mouse-brain}. Other theoretical approaches to investigate myelin sheath light guidance utilise an idealized representation of the myelinated axon to obtain electromagnetic field solutions. For this simple model in the context of the optical and infrared spectral range, it has been demonstrated analytically that localized weakly damped modes can exist in myelinated axons\cite{analytical-myelinated-axon}. It has also been recognized that the cross-section of myelinated nerve fiber can be considered a depressed-core fiber due to its refractive index profile. Myelinated nerve fiber can be treated as an anti-resonant reflecting optical waveguide\cite{ARROW-nerve-fiber-simulation} (a model developed for depressed-core fibers\cite{DCF-ARROW-theory}) wherein myelin, as the high-index cladding, will confine and transmit incident light well even in the presence of optical imperfections. \\
\\
There are a number of experimental studies that support the plausibility of light guidance along myelinated axons. In particular, a recent study\cite{white-matter-tracts-better} demonstrates a clear directional dependence on light transmission across myelinated axons, finding a 50\% decrease in scattering power for light propagation along highly myelinated and organized (spinal cord) white matter tracts. In contrast, directional dependence in overall light attenuation did not occur in grey matter where there is very little myelin and myelin organization. Images from this experimental study also show increased transmission through the myelin sheath compared to the axon, further highlighting the importance of the myelin sheath to light guidance along nerve fibers. These findings confirm prior indirect evidence\cite{white-matter-tracts-old} for light conduction along white matter tracts. Other experimental studies have also observed increased biophoton activity at the end of nerve roots in response to light stimulation\cite{biophoton-neural-signals} and at axons and axon terminals in response to glutamate stimulation\cite{spatiotemporal-glutamate-imaging}. \\
\\
Photonic communication in the brain would have unique features due to the nature of biophotons\cite{biophoton-brain-signal-review, group-review}. While biophoton signal speed would depend on emission and detection rates, photons travel at the speed of light, and therefore, such signals could potentially be very fast relative to the electrochemical processes for action potentials. Furthermore, biophotonic signals could communicate information in multiple ways -- by photon number, frequency, and polarization. Since photon polarization is well suited for encoding quantum information, biophoton signals open the possibility for quantum communication in the brain\cite{Sourabh, group-review, simon-consciousness}. So, it seems pertinent to assess the effective transmission of biophotons and their polarization in the brain environment as a biologically realistic phenomenon. \\
\\
We study how a guided mode propagates through the structure of the myelinated axon by observing polarization evolution and transmission through multiple Ranvier nodes. This is the first work considering the evolution of light polarization in the myelin sheath. In Methods, we describe our myelinated axon model (which is informed by biological observations) and details of the software as they pertain to our findings. In Results, we present observations of polarization evolution and transmission of a guided mode through the myelinated axon model with four Ranvier nodes. The significance of the results suggested improvements to the model, and applications are discussed. We also briefly present calculations of other modes of the myelin sheath in Appendix A. Additional figures to observe polarization evolution in the myelin sheath are included in Supplemental Information.

\section*{Methods}
This section details the construction of our myelinated axon model and discusses relevant software details and settings as they pertain to our results. ANSYS Lumerical FDTD: 3D Electromagnetic Simulator software is used for model construction and simulation of light propagation.

\subsection*{Model design}

\subsubsection*{Input light and material optical properties}
We specify the wavelength as $\lambda = 0.4 \: \mu \text{m}$ to simulate biophotons as the light source. This falls within the biophotonic spectral range\cite{UWPE-review} of $0.35 \: \mu \text{m}$ -- $1.3 \: \mu \text{m}$. We choose to restrict our input light to a single (short) wavelength within this range to offset the computational demands of the simulation software. Shorter wavelengths can be well-confined in myelinated axons with thinner myelin with correspondingly smaller cross-sectional dimensions, decreasing the simulation region volume required to contain the model. \\
\\
We consider the myelinated axon composed of three fundamental materials: axon, myelin sheath, and interstitial fluid. These are all modelled as uniform dielectric materials of specified refractive index. We set refractive indices to $n_{a} = 1.38$ for the axon\cite{axon-refractive-index}, $n_{m} = 1.44$ for the myelin sheath\cite{myelin-refractive-index}, and $n_{int} = 1.34$ for the interstitial fluid\cite{fluid-refractive-index} as consistent with biological observations. This refractive index profile highlights the high refractive index of the myelin sheath, which underpins its waveguiding ability. \\
\\
In our modeling, we assume that our choice of wavelength, and the larger biophoton spectral range, face negligible absorption by the myelin sheath over inter-neuron distances on the centimeter scale as previously discussed\cite{Sourabh}. The myelin sheath is composed of lipids, proteins, and water. In consideration of lipids, we recognize that mammalian fat has a very low absorption coefficient (less than $0.01 \: \text{mm}^{-1}$) for the biophoton spectral range\cite{mammal-fat-absorbs}. Water has a similarly low absorption coefficient. Similar to those found in the myelin sheath, most proteins have negligible absorption for wavelengths around $0.34 \: \mu \text{m}$ and above\cite{protein-absorbs}. The assumption of low biophoton absorption in the myelin sheath is also indirectly supported by studies on brain tissue. Absorption in white matter is relatively low (its absorption coefficient decreasing almost monotonically from $\sim 0.3 \: \text{mm}^{-1}$ to $\sim 0.07 \: \text{mm}^{-1}$ over the wavelength range of $0.4 \: \mu \text{m}$ -- $1.1 \: \mu \text{m}$), wherein the presence of myelin is thought to have little effect as the grey matter has comparable absorption coefficients\cite{white-grey-matter-absorbs}. Observations have also shown that optical attenuation in brain tissue is primarily due to scattering and not absorption\cite{brain-scatters-not-absorbs}.  

\subsubsection*{Myelinated axon structure}
\begin{figure}[ht]
\centering
\includegraphics[scale = 0.35]{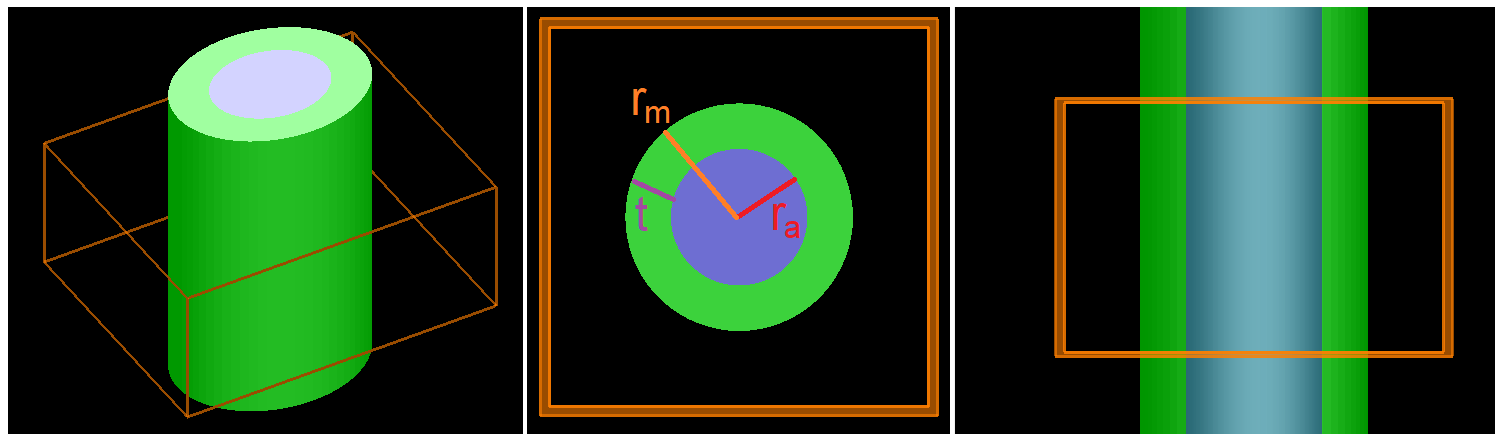}
  \caption{Internodal region of myelinated axon model. The blue area represents the axon, the green area represents the myelin sheath, and the orange lines are included to enhance perspective and depict a test simulation region. The leftmost image depicts the perspective view, center image depicts the cross-section, and the rightmost image depicts the side view. The cross-section of the internodal region is annotated with the axon radius $r_{a}$, the outer myelin radius $r_{m}$, and the myelin thickness $t$. Images taken from ANSYS Lumerical FDTD: 3D Electromagnetic Simulator software.}\label{img:basic-model}
\end{figure}

\FloatBarrier
The biological axon (also referred to as a nerve fiber) is roughly cylindrical in shape, located in an interstitial fluid environment. We model the myelinated axon with an ideal cylindrical structure with specified dimensions. We take the axis symmetry to be the $z$ axis, which coincides with the propagation direction of the input light. The only $z$ dependence in the model is due to the Ranvier nodes, the myelin sheath being present in the internodal regions and absent in the nodal regions. \\
\\
Internodal regions feature the axon and surrounding myelin sheath as illustrated from multiple perspectives in Fig. \ref{img:basic-model}, for which we now describe the cross-section. An inner cylinder of radius $r_{a}$ models the axon. An outer cylindrical ring of inner radius $r_{a}$ and outer radius $r_{m}$ models the myelin sheath. The ratio between $r_{a}$ and $r_{m}$ is known as the g-ratio, such that $r_{a} = g \cdot r_{m}$. We take $g = 0.6$ which is close to the experimental average\cite{internodal-study}. As we have chosen $\lambda = 0.4 \: \mu \text{m}$ for our input light, we set the myelin thickness $t = r_{m} - r_{a} = 0.4 \: \mu \text{m}$. This ensures good confinement for the light within the myelin sheath while ensuring manageable computational demands. Taken together with the g-ratio, this corresponds to an axon radius of $r_{a} = 0.6 \: \mu \text{m}$. This is consistent with the biologically observed values of axon diameter, which are between 0.2 $\mu \text{m}$ – 10 $\mu \text{m}$ in cortical white matter\cite{axon-diameters}. So, we set axon radius $r_{a} = 0.6 \: \mu \text{m}$ and myelin radius $r_{m} = 1 \: \mu \text{m}$. \\
\\
Ranvier nodes, segments of the bare axons with no myelin sheath, interrupt internodal regions at regular intervals. They are a critical feature to consider when assessing the ability of the myelin sheath to serve as a waveguide since this presents a significant opportunity for transmission loss. The Ranvier node is known to be roughly $1 \: \mu \text{m}$ - $2 \: \mu \text{m}$ in length\cite{node-length-affects-signal-speed, book-ranvier-node}. In our model, we set Ranvier node length to be $2 \: \mu \text{m}$, such that the myelin sheath abruptly ends at these boundaries and its cross-section only includes that of the axon. In this work, in order to limit the complexity of our multi-node calculations, we do not attempt to model the paranodal regions located at the boundaries between the internode and Ranvier node segments at which the constituent cytoplasmic loops of the myelin taper off onto the axon. There is significant variation in internode distance within the central nervous system; however, internode distance has a strong relationship with the outer myelin diameter of the nerve fiber\cite{internodal-study}. It has been observed that fibers with an outer myelin diameter of $2 \: \mu \text{m}$ (as in our case) roughly correspond to $100 \: \mu \text{m}$ in internode distances\cite{internodal-study}. In our model, we set the node-to-node distance to be $100 \: \mu \text{m}$, such that the Ranvier node is 2 $\mu \text{m}$ in length and the internode is 98 $\mu \text{m}$ in length. Our myelinated axon model is $500 \: \mu \text{m}$ in length and includes four Ranvier nodes placed at $z = 100, 200, 300, 400 \: \mu \text{m}$, as shown later in Fig. \ref{fig:four-node-structure}. 

\subsection*{Software: Simulation Region and Monitors}
This project utilised ANSYS Lumerical FDTD: 3D Electromagnetic Simulator software (version: 8.26.2779) to construct models and run simulations. This software uses the Finite Difference Time Domain (FDTD) method to solve Maxwell’s equations for models on the nanometer scale, which allows us to calculate modes and simulate light propagation in our myelinated axon model. The FDTD method solves the electric and magnetic field components on a discrete spatial and temporal grid within the Yee cell paradigm. \\
\\
We consistently set the simulation region to span $4 \: \mu \text{m}$ in the $x$ and $y$ directions so that the cross-section is square. These dimensions were chosen to generously allow more than a wavelength distance ($\lambda = 0.4 \: \mu \text{m}$) between the outer myelin radius and the simulation boundary, beyond which we say light is lost to the external environment. The simulation region spans $500 \: \mu \text{m}$ in the $z$ direction. The input mode is applied as a planar source placed at the beginning of the region at $z = 0 \: \mu \text{m}$. The myelinated axon model is always centered in the simulation region cross-section and uses the entire region length to consider light propagation through multiple Ranvier nodes. All boundaries of this simulation region are set to PML (perfectly matched layer), which models perfect absorption and simulates complete light loss at the boundaries. \\
\\
In this work, we collect transmission and electric field profile data at twenty cross-sectional monitors throughout the $500 \: \mu \text{m}$ length of the simulation region, which contains the myelinated axon structure having four Ranvier nodes. Monitors were located along the $z$ axis with the same $x$ and $y$ boundaries as the simulation region (each spanning $4 \: \mu \text{m}$). They were placed $25 \: \mu \text{m}$ apart, starting at $z = 25\: \mu \text{m}$ and ending at $z = 500 \: \mu \text{m}$. This was taken to be adequate sampling to observe changes in the electric field, being significantly smaller than the internodal length of $98 \: \mu \text{m}$ (as Ranvier nodes cause changes in the propagation of the light). The software automatically snaps these monitors to the location of the nearest mesh cell, designed to improve data accuracy by minimizing the interpolation required. Monitors placed at the boundaries of the nodes and at the boundary at the end of the simulation region may have some slight inaccuracies due to this interpolation. 

\section*{Results}
\begin{figure}[ht]
\centering
\includegraphics[width = \textwidth]{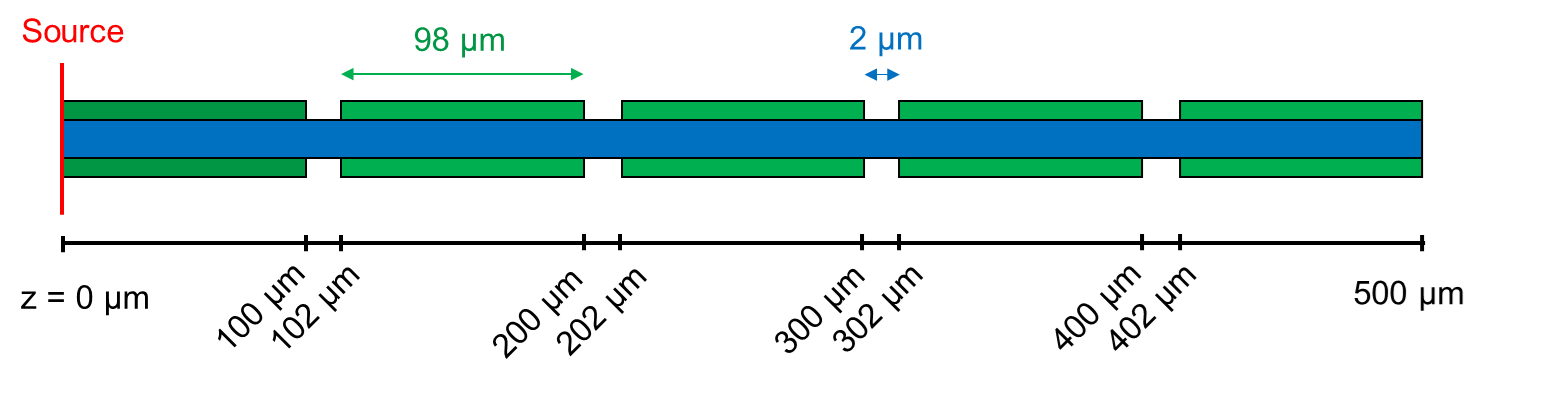}
\caption{Longitudinal diagram of the four-node structure. The blue area represents the axon, and the green areas represent the myelin sheath. The nodes are 2 $\mu \text{m}$ in width and are spaced 100 $\mu \text{m}$ apart. Note that $z = 0$ $\mu \text{m}$ is the beginning of the simulation and location of the mode source, $z = 500$ $\mu \text{m}$ is the end of the simulation.} \label{fig:four-node-structure}
\end{figure}

\begin{figure}[ht]
\centering
\includegraphics[width=1\linewidth]{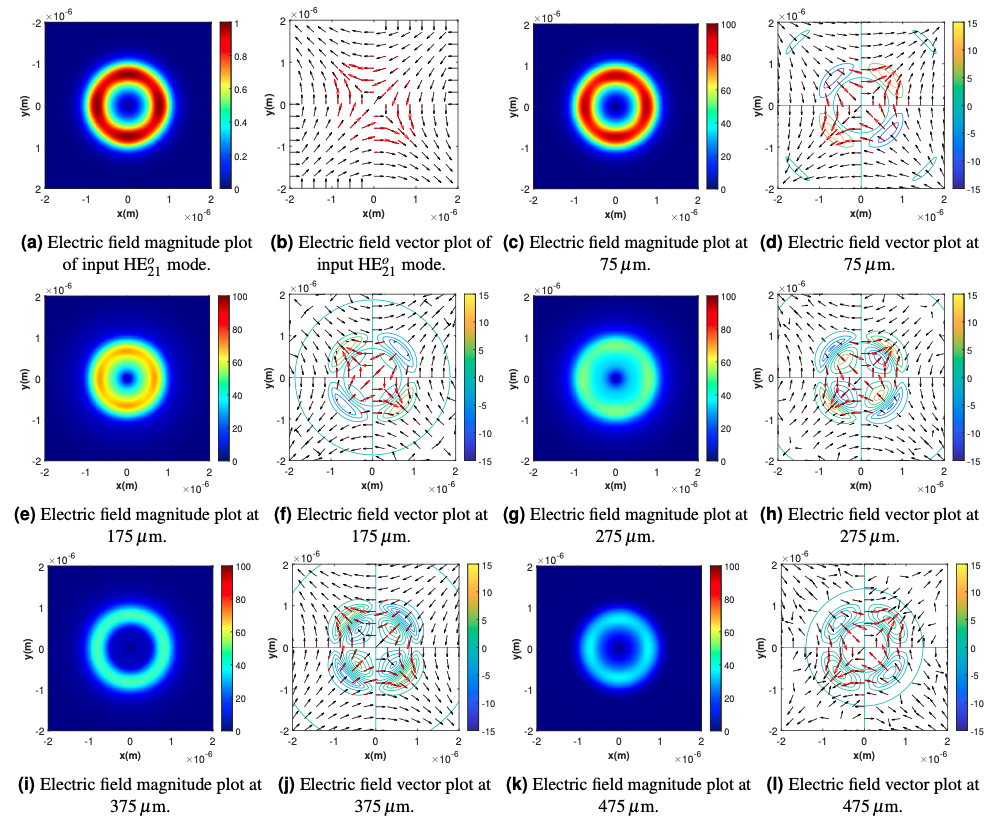}
  \caption{Electric field profiles of calculated input mode HE$_{21}^{o}$ and cross-sections of the four-Ranvier-node myelinated axon structure at positions $75, 175, 275, 375, 475 \: \mu \text{m}$. The mode source is located at $0 \: \mu \text{m}$, input amplitude is $100 \: \text{V}/\text{m}$. Ranvier nodes are $2 \: \mu \text{m}$ long and begin at $z = 100, 200, 300, 400 \: \mu \text{m}$. Plots (a), (c), (e), (g), (i), (k) depict electric field magnitude. Plots (b), (d), (f), (h), (j), (l) depict electric field vectors; black arrows depict the normalized three-dimensional electric field vector at a given position, red arrows depict the magnitude-scaled three-dimensional electric field vector at a given position; color-coded contour lines depict the magnitude of $E_z$. Electric field magnitudes given in units of $\text{V}/\text{m}$.}\label{fig:electric-field-profiles}
\end{figure}

\begin{figure}[ht]
\centering
\includegraphics[width=0.75\linewidth]{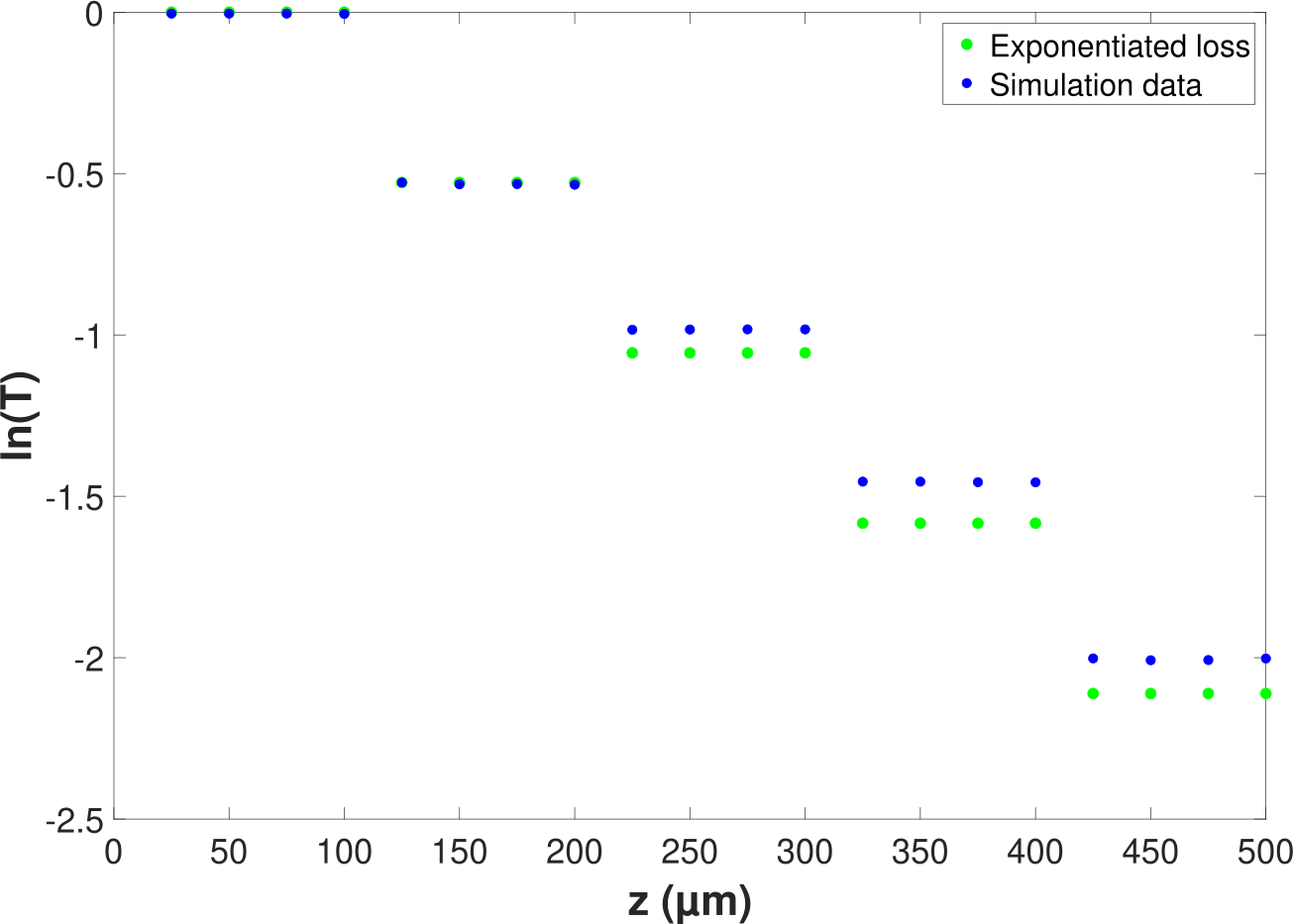}
  \caption{Natural log of transmission coefficient $T$ as a function of distance $z$ along the four-Ranvier-node myelinated axon structure. The blue points show transmission data collected from the simulation; the green points show transmission data expected from the assumption of exponentiated loss through each Ranvier node, using the measured transmission loss from the first Ranvier node.  The mode source is located at $z = 0 \: \mu \text{m}$, Ranvier nodes are $2 \: \mu \text{m}$ long and begin at $z = 100, 200, 300, 400 \: \mu \text{m}$.}\label{fig:transmission}
\end{figure}

\begin{figure}[ht]
\centering
\includegraphics[width=0.75\linewidth]{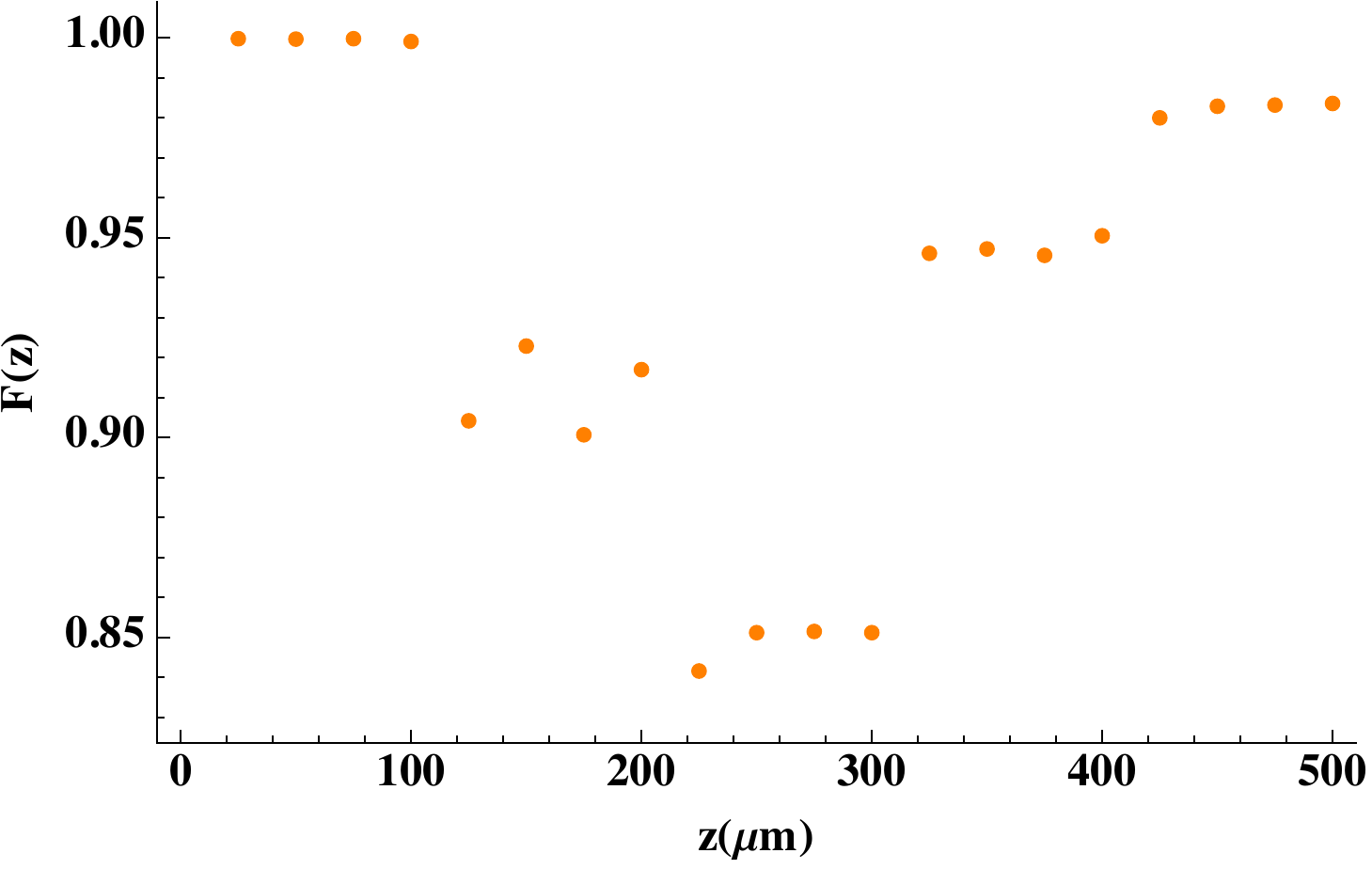}
  \caption{Polarization transmission fidelity $F(z)$, defined in Eq. \ref{eq:over}, as a function of longitudinal position $z$. }\label{fig:over}
\end{figure}

\FloatBarrier
In this section, we simulate the propagation of a mode through the myelinated axon structure with four Ranvier nodes. This allows us to observe the electric field profile, both its magnitude and polarization, at different cross-sections throughout the structure. We present the input mode and the electric field profiles at $z = 75, 175, 275, 375, 475 \: \mu \text{m}$ in Fig. \ref{fig:electric-field-profiles}, which shows electric field behavior well after any given Ranvier node. Electric field profiles at a resolution of $25 \: \mu \text{m}$ are presented in Supplemental Information. Simulating light propagation here also allows us to characterize the transmission drop-off through multiple Ranvier nodes. Transmission as a function of distance along the myelin sheath is presented in a log plot in Fig. \ref{fig:transmission}. Overlap of the complex electric field cross-sections with the input mode is presented as a function of distance along the myelin sheath in Fig. \ref{fig:over}.\\
\\
The modes of the myelinated axon structure have been calculated and examined in previous studies\cite{biophoton-propagation-simulation}. These are supported by mode calculations for our model. We present the first six modes calculated for our myelinated axon model in Appendix A to further highlight various polarization patterns of guided light in the ideal myelin sheath. The HE$_{21}^{o}$ mode, shown in Fig. \ref{fig:electric-field-profiles}a -- \ref{fig:electric-field-profiles}b, is used as the input mode for our four-Ranvier-node structure. The magnitude plot of this mode in Fig. \ref{fig:electric-field-profiles}a shows four high-magnitude areas we will refer to as anti-nodes, which are equally spaced and interspersed with low-magnitude nodes. The vector plot in Fig. \ref{fig:electric-field-profiles}b shows its hyperbolic polarization pattern. Input light amplitude is set to $100 \: \text{V}/\text{m}$. \\
\\
Our four-Ranvier-node myelinated axon structure is $500 \: \mu \text{m}$ long as shown in Fig. \ref{fig:four-node-structure}, measured along the $z$ axis which is the symmetry axis for the model. Input light is applied at $z = 0 \: \mu \text{m}$. Ranvier nodes are taken to be $2 \: \mu \text{m}$ gaps in the myelin sheath, located at $z = 100, 200, 300, 400 \: \mu \text{m}$. \\
\\
\\
\noindent At $75 \: \mu \text{m}$ in the myelinated axon structure, the electric field profile is shown in Fig. \ref{fig:electric-field-profiles}c -- \ref{fig:electric-field-profiles}d. This magnitude and vector plot at $75 \: \mu \text{m}$ match those of the mode calculated at $0 \: \mu \text{m}$. As expected, the mode remains confined to the myelin sheath, and there is no visually obvious change in its polarization. (The only noticeable change in the magnitude plots is the nonzero $E_z$ component at $75 \: \mu \text{m}$, in contrast to the calculated mode at $0 \: \mu \text{m}$, due to the mode's forward propagation along the $z$ axis.) It is also propagating with 100\% transmission as expected, indicated in the transmission plot in Fig. \ref{fig:transmission} prior to the first node at $100 \: \mu \text{m}$. At $175 \: \mu \text{m}$, after the first Ranvier node, the electric field profile is shown in Fig. \ref{fig:electric-field-profiles}e -- \ref{fig:electric-field-profiles}f. There is clearly a decrease in magnitude due to the Ranvier node; however, we still see the light confined to the myelin sheath and four high-magnitude anti-nodes. The hyperbolic polarization pattern still appears well-preserved in the myelin sheath. At $275 \: \mu \text{m}$, after two Ranvier nodes, the electric field profile is shown in Fig. \ref{fig:electric-field-profiles}g -- \ref{fig:electric-field-profiles}h. At this position, the magnitude has been further reduced from the Ranvier nodes, and we see more light propagating in the axon. The polarization still appears well-preserved. At $375 \: \mu \text{m}$ (after three Ranvier nodes) shown in Fig. \ref{fig:electric-field-profiles}i -- \ref{fig:electric-field-profiles}j, and at $475 \: \mu \text{m}$ (after four Ranvier nodes) shown in Fig. \ref{fig:electric-field-profiles}k -- \ref{fig:electric-field-profiles}l, we see similar trends. The magnitude plot evolves according to position, with magnitudes generally decreasing due to loss caused by the Ranvier nodes but continuing to propagate within the myelin sheath. The hyperbolic polarization pattern appears to stay well-preserved within the myelin sheath, where there is still significant light transmission. \\

\noindent The transmission coefficient $T$ was recorded every $25 \: \mu \text{m}$ along the $z$ length of the myelinated axon structure for the cross-sectional area of the simulation region ($4 \: \mu \text{m}$ x $4 \: \mu \text{m}$). The natural log of $T$ is plotted against $z$ in Fig. \ref{fig:transmission}, plotting recorded simulation data in blue. We see perfect transmission prior to $100 \: \mu \text{m}$, meaning before the first Ranvier node, as expected from the input mode of the model. The four decreases in transmission occur at sample cross-sections directly after the four Ranvier nodes at $z = 100, 200, 300, 400 \: \mu \text{m}$. We compare this data to the prior assumption that the transmission across successive Ranvier nodes drops off according to exponentiated transmission loss. The transmission loss plotted in green depicts the expected transmission loss with this assumption, using the transmission coefficient after the first Ranvier node ($T = 0.5899$ measured at $z = 125 \: \mu \text{m}$). With this assumption, our simulation results show slightly higher transmission through subsequent Ranvier nodes than expected. 

To further study the polarization transmission, we calculated the normalized overlap of the complex electric field profile at various points with the input field profiles, shown in Fig. \ref{fig:over}. Specifically, we define the polarization transmission fidelity $F(z)$ as a function of longitudinal position $z$ as follows:
\begin{ceqn}
\begin{align}
 \begin{gathered}
F(z) = \frac{|\int \int \langle \vec{E}(x,y,0) , \; \vec{E}(x,y,z) \rangle  dx dy|}{\sqrt{\int \int  \langle \vec{E}(x,y,0) , \; \vec{E}(x,y,0) \rangle  dx dy} \sqrt{\int \int  \langle \vec{E}(x,y,z) , \; \vec{E}(x,y,z) \rangle dx dy}}
 \end{gathered}
 \label{eq:over}
\end{align}
\end{ceqn}
 where $\langle ... ,\; ... \rangle $ is the inner product of two complex vectors. Fig. \ref{fig:over} shows some oscillations that are likely due to the mixing of lossy modes at the Ranvier nodes, which then interfere with the propagating mode. However, as expected from visual inspection of the mode profiles in Fig. \ref{fig:electric-field-profiles}, the fidelity generally remains quite high at all points. In particular, it returns to values very close to one after the fourth node.

\FloatBarrier

\section*{Discussion}
We have investigated the properties of guided light propagation in the myelin sheath with multiple Ranvier nodes, utilising ANSYS Lumerical FDTD: 3D Electromagnetic Simulator. Our myelinated axon model was constructed according to biologically realistic structural and optical properties for axon, myelin sheath, and interstitial fluid material. We observed polarization evolution and transmission through multiple Ranvier nodes. The input mode's polarization state is well-preserved through the four-Ranvier-node myelin sheath model, which we have confirmed by calculating the normalized overlap between the electric field profiles in the model and the input mode itself. We observe high polarization fidelity between the electric field and the input mode throughout the myelinated axon model, with some oscillations that are likely due to interference with lossy modes produced by the Ranvier nodes. Furthermore, while it has been previously assumed\cite{Sourabh} that light travelling through multiple Ranvier nodes may obey exponentiated transmission loss, results from our model show slightly higher transmission than this expectation. \\
\\
There are a number of limitations for this kind of model in its ability to describe potential biophoton travel through the myelin sheath. For input light, we use a single mode, which does not consider the point-like nature of the localized character of biophoton sources, nor does it study how biophotons couple into the modes of the myelin sheath. Possible biophoton sources in neurons include mitochondria\cite{mitochondria-source-observation, mitochondria-source-theoretical} (mitochondrial respiration) and liposomes\cite{liposomes-source} (lipid oxidation). Mitochondria are particularly interesting as a potential source of biophotons in the context of a recent study\cite{mitochondria-decode-firing} which indicates mitochondria are capable of encoding neuron firing frequency, as this could be a potential connection between action potential signals and biophoton signals. For light detectors, our study uses planar monitors to record electric field data for cross-sections of the myelinated axon model; however, we would also expect biological biophoton detectors to be localized in nature. Possible light-sensitive detectors in neurons include centrosomes\cite{centrosomes-detector} and chromophores within mitochondria\cite{mitochondria-cytochrome-detector}. Brain tissue is known to be light-sensitive in general\cite{light-brain-cortical-tissue, light-brain-NMDA, light-brain-modulates-cognition, light-brain-alters-connectivity}, even containing light-sensitive opsin molecules within deep brain tissue\cite{light-brain-alters-opsins}. \\
\\
There are a number of features that can be added to our myelinated axon model to better describe the biological reality\cite{Sourabh, biophoton-propagation-simulation, ARROW-nerve-fiber-simulation}, as has been examined in previous studies thus far to measure transmission through short segments of the myelin sheath. Polarization evolution in the myelin sheath can be assessed across realistic ranges of biological parameters -- notably axon radius, g-ratio, and biophoton wavelength -- and in the presence of realistic optical imperfections such as bends in the axon, varying myelin thickness, irregular cross-section shapes and areas, paranodal regions and the positive radial birefringence of the myelin sheath\cite{biref-myelin-review, biref-myelin-measured}. \\
\\
Understanding how polarization evolves as light travels through the myelin sheath can help us determine if information encoded within polarization can be transmitted through these waveguides between neurons. If so, this allows the possibility of biophotonic quantum communication, as photon polarization is able to encode quantum information (in qubits) in addition to expected classical information (in bits). There is growing evidence that quantum effects may play significant roles in biological environments, including the brain\cite{quantum-in-brain}. It has been suggested\cite{simon-consciousness} that the exchange of quantum information via biophotons could potentially sustain an entangled system of spins within neurons in the brain. Quantum computation in the brain\cite{quantum-cognition} might provide a performance advantage, which is well-demonstrated in quantum computing for certain types of problems\cite{https://doi.org/10.48550/arxiv.2203.17181}. Such a quantum advantage might be beneficial to meet the high computational demands of the brain under energy constraints within the body, especially for functions such as consciousness. Quantum entanglement may also play a role in consciousness, with the potential to explain its unified and complex nature\cite{original-quantum-consciousness, simon-consciousness}. So, understanding the nature of biophotons as possible information carriers, either classical or quantum in nature, may have many significant applications within the field of neuroscience. \\
\\
There are a number of other motivations to characterize the optical properties of the myelin sheath outside of possible biophotonic communication. The myelin sheath is also examined as a waveguide within the theory that Nodes of Ranvier act as relay amplifiers and produce infrared photons to facilitate saltatory conduction along myelinated axons\cite{bio-nanoantennas, softmaterial-waveguides, infrared-simulation, infrared-in-vivo-study}. This is a separate topic from guided biophotons along myelinated axons, considering a different spectral range with Ranvier nodes as sources. It is also useful to characterize light interactions with the myelin sheath for the purposes of neuroimaging. Application of polarized light to brain tissue allows for 3D-PLI (3D polarized light imaging) \cite{polarized-neuroimaging, jones-matrix-paper} to image nerve fibers at a microscopic level. Entangled light might also be useful for imaging purposes\cite{entangled-light-imaging-alzheimers, entangled-light-imaging-mouse-brain}. It is important to consider how light interacts with neural cells in general, for applications in nanoscale optogenetics\cite{nanoscale-optogenetics}. From a medical perspective, investigation of optical properties of the myelin sheath may lead to a greater understanding of causes and potential treatments of diseases associated with it (such as multiple sclerosis\cite{MS-disease}).

\bibliography{sample}



\section*{Acknowledgements}

The authors would like to thank Paul Barclay for helpful discussions. This work was supported by an NSERC Discovery Grant and by NRC CSTIP Grant QSP 022.

\section*{Author contributions statement}

E.F. performed the calculations with help from H.Z.-H. and wrote the paper with help from H.Z.-H. and C.S.  C.S. conceived and supervised the project.\\
\\

\section*{Data availability}
The datasets of this study are available on \href{https://github.com/hadizh20/ResearchProj_MyelinSheathWaveguides}{Github}.





\appendix

\section{Myelin sheath modes}
\FloatBarrier
We examine the guided modes for the myelinated axon model. As described in a previous study simulating optical modes in the myelin sheath\cite{biophoton-propagation-simulation}, guided modes obey the eigenmode wave equation as shown in Equation \ref{eq:mode-eq} for the boundary conditions applied by its waveguide, in this case, the myelin sheath. As laid out in the previous investigation of myelin sheath modes\cite{biophoton-propagation-simulation} considering myelin sheath modes, the solution of the wave equation $\vec{E}$ and its eigenvalue $\lambda_{e}$  are as stated in Equation \ref{eq:mode-sol} -- where $z$ is the propagation axis, $\omega$ the angular frequency, $\epsilon$ the relative dielectric constant, $\mu$ the relative magnetic permeability, and $k_{z}$ the longitudinal propagation constant.

\begin{ceqn}
\begin{align}
 \begin{gathered}
\nabla^2 \vec{E} \left( \vec{r}, t \right) + \lambda_{e} \vec{E} \left( \vec{r}, t \right) = 0 \\
    \nabla^2 \vec{H} \left( \vec{r}, t \right) + \lambda_{e} \vec{H} \left( \vec{r}, t \right) = 0 
 \end{gathered}
 \label{eq:mode-eq}
\end{align}
\end{ceqn}

\begin{figure}[ht]
\centering
\includegraphics[width=1\linewidth]{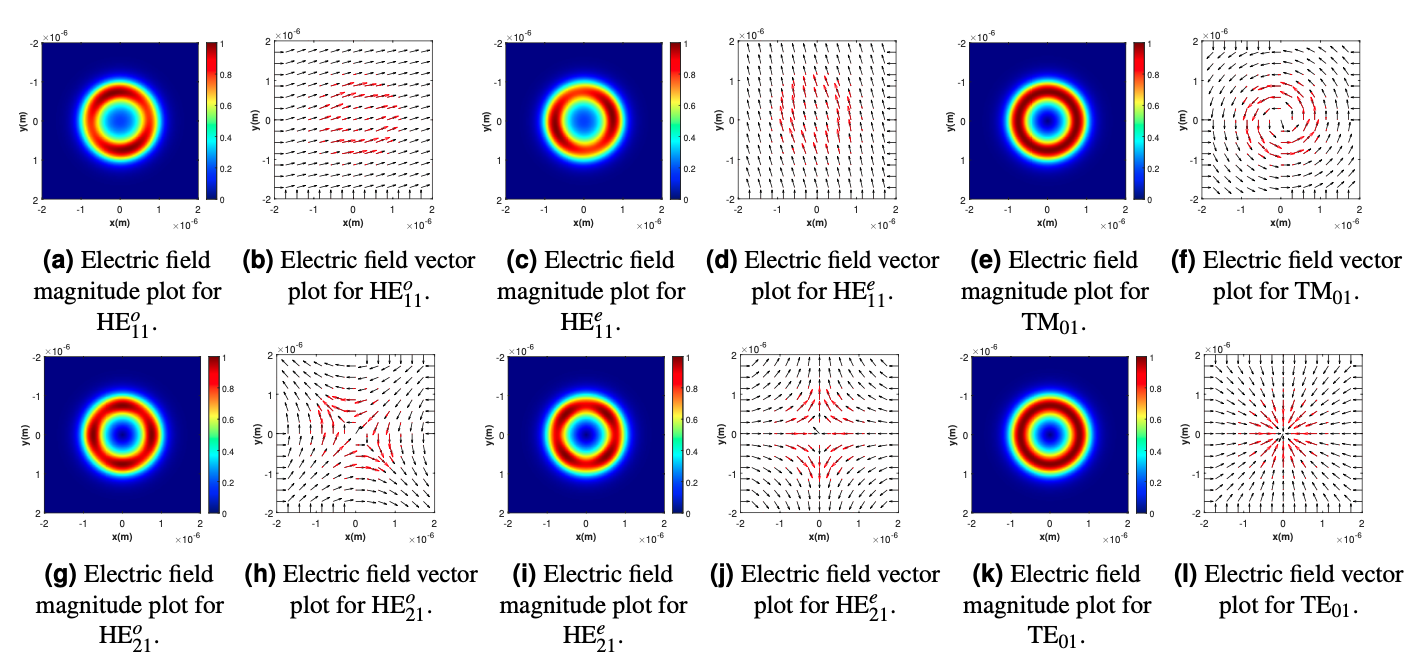}
  \caption{The first six modes calculated by the software (as ordered by decreasing effective index), named according to established convention\cite{biophoton-propagation-simulation, modes-book, modes-paper}. Modes are calculated over the cross-section of the model at the source ($z=0 \: \mu \text{m}$). Input light amplitude is $1 \: \text{V}/\text{m}$. Plots (a), (c), (e), (g), (i), (k) depict electric field magnitude. Electric field magnitudes given in units of $\text{V}/\text{m}$. Plots (b), (d), (f), (h), (j), (l) depict electric field vectors; black arrows depict the normalized three-dimensional electric field vector at a given position, red arrows depict the magnitude-scaled three-dimensional electric field vector at a given position. The $E_z$ component is negligible in all cases, on the order of $10^{-17} \: \text{V}/\text{m}$ or less.}\label{fig:all-modes}
\end{figure}

\begin{ceqn}
\begin{align}
 \begin{gathered}
 \vec{E} \left( \vec{r}, t \right) = \vec{E} \left( x, y \right) e^{i \left( \omega t - k_{z} t \right) } \\
        \lambda_{e} = \omega^2 \epsilon \mu - k_{z}^2
 \end{gathered}
 \label{eq:mode-sol}
\end{align}
\end{ceqn}
The propagation constant $k_z $  may also be expressed as the effective index $n_{eff} = k_z / k_0$, where $k_0$ is the vacuum propagation constant. We calculate six modes of the cross-section of our myelinated axon structure, ordered according to decreasing effective index. Electric field magnitude and vector plots are shown in Fig. \ref{fig:all-modes}. Modes are labelled according to the established conventions for circular isotropic optical fibers\cite{biophoton-propagation-simulation, modes-book, modes-paper}. For each mode, we also record and discuss its effective index, which is analogous to its eigenvalue. For all modes, we see that the electric field is well-confined to the myelin sheath (centered on the origin, inner radius of $0.6 \: \mu \text{m}$, outer radius of $1 \: \mu \text{m}$). \\
\\
Most of the modes have magnitude plots which alternate between brighter, high-magnitude areas we refer to as anti-nodes, and dimmer, lower-magnitude areas, which we will refer to as nodes. These modes do not have cylindrical symmetry, however, this is expected as they are degenerate (sharing an effective index) and can thus be sorted into pairs related by a simple rotation. HE$_{11}^{o}$ (Fig. 6a – 6b) and HE$_{11}^{e}$ (Fig. 6c – 6d) have roughly linear polarization patterns and an effective index of $n_{eff} = 1.4412$. They each have two anti-nodes and two nodes, which are related by a 90-degree rotation. HE$_{21}^{o}$ (Fig. 6g – 6h) and HE$_{21}^{e}$ (Fig. 6i – 6j) have roughly hyperbolic polarization patterns and an effective index of $n_{eff} = 1.4082$. They each have four anti-nodes and four nodes, which are related by a 45-degree rotation. We also observe two modes with cylindrical symmetry which are non-degenerate. They have magnitudes equally distributed through the ring of the myelin sheath without any anti-nodes or nodes. TM$_{01}$ (Fig. 6e – 6f) has azimuthal polarization and an effective index of $n_{eff} = 1.4090$. TE$_{01}$ (Fig. 6k – 6l) has radial polarization and an effective index of $n_{eff} = 1.4071$.

\end{document}